\documentclass{revtex4}

\usepackage{natbib}
\usepackage{graphicx}

\begin{document}

\title{Use of a Microcontroller for Fast Feedback Control of a Fiber Laser}

\author{M. R. Dietrich}
\email{dietricm@u.washington.edu}
\author{B. B. Blinov}
\affiliation{University of Washington Department of Physics, Seattle, Washington, 98195}

\begin{abstract}
An inexpensive,  easily programmed microcontroller is demonstrated for the fast frequency stabilization of an infrared fiber laser.  The microcontroller manages all digitalization and processing, with external circuitry only providing buffering, amplification and overvoltage protection.  Several convenient features are included to recover from common failure modes of a laser lock system, and a final Nyquist frequency of 200 kHz is obtained, corresponding to a usable bandwidth of 40-60 kHz.
\end{abstract}

\maketitle

\section{Motivation}

A common problem addressed in atomic physics laboratories is the need to stabilize the frequency of a laser with respect to some reference, such as a resonance in some molecular species or optical cavity, which is traditionally realized by an analog proportional-integral-derivative (PID) feedback system.  Although such a circuit is simple enough on its own, certain other desirable features such as an available scanning mode, unlock detection, and automatic lock reacquisition can quickly make an analog design cumbersome.  The construction and maintenance of these systems can consume great quantities of lab worker time.  With a digital design, however, these objectives can all be accomplished in software, enabling rapid development and modification.  By choosing a commercial microcontroller containing almost all necessary circuit components such as analog to digital converters (ADC), digital to analog converters (DAC), and timers the design can also be quickly reproduced for additional applications.  The use of a microcontroller as the master director of highly elaborate, multi-component laser stabilization schemes has been demonstrated before for He-Ne lasers\cite{Lazar1997, Ahola1998, Eom2002}, and external cavity diode lasers (ECDL)\cite{Allard2004}.  Here we discuss the simpler task of direct laser frequency control through high bandwidth feedback.

Our target system was a Koheras A/S Adjustik\texttrademark~fiber laser which was to be stabilized to an optical cavity by the Pound-Drever-Hall (PDH) technique\cite{Black2001}.  Part of the objective was to achieve line narrowing, and so high bandwidth of the feedback circuit was desirable.  Also, previous experience with the system had demonstrated a tendency for the laser to unlock and relock on an adjacent cavity line.  Therefore, we sought a high bandwidth system that would detect an unlock condition and recapture on the original line before locking on the next.  To achieve this, we settled on a design having two modes of operation.  In the scanning mode, the circuit would scan over a configurable range, which would be set by the operator to include only a single line.  In the lock mode, the circuit would lock to lines only within the previously set scanning range.  If ever during lock mode operation the output reached either rail of the set scanning range, the circuit would gently return to the opposite rail and restart the search.

\section{System Description}

\begin{figure}
\includegraphics[width=6in]{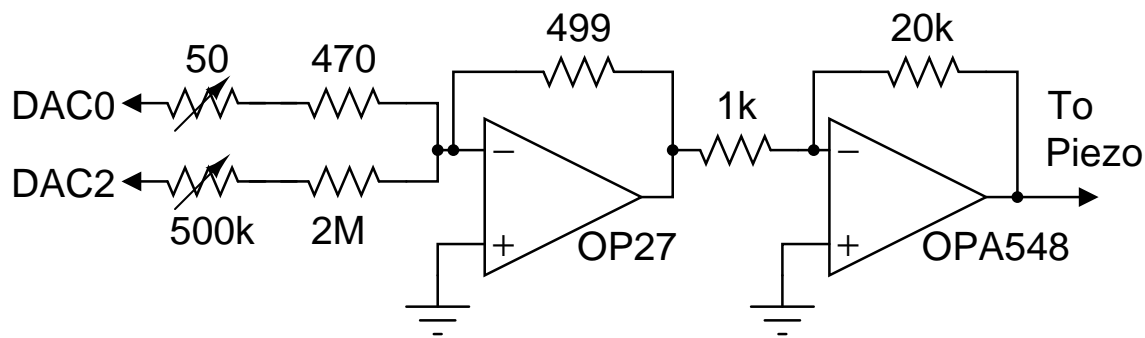}
\caption{Circuit diagram of the DAC combiner circuit and output amplification stage.  The resistor values are chosen to allow fine tuning with the two potentiometers, which we set to ensure that DAC0 is weighted $2^{12}$ times more strongly than DAC2.  DAC1 is avoided because the minikit connects it to an LED.}
\label{dacout}
\end{figure}

To realize these features, we chose an ADuC7020 microcontroller from Analog Devices.  This system has built in five 12-bit ADC's and four 12-bit DAC's, all capable of sampling at a rate of 1 MHz.  An inexpensive development kit is available (``Minikit'', EVAL-ADUC7020MKZ), which is easily soldered to standard circuit boards.  One ADC was dedicated to measuring the error signal from the PDH system while the other four monitored potentiometers, which are the user controls.  Two potentiometers set the lower rail and sweep width, while the other two set the proportional and integral gain parameters.  Because our lock bandwidth is limited by noise, rather than the laser piezo's resonance, no differential gain was implemented or felt necessary.  It was decided that the dynamic range offered by a 12-bit DAC was insufficient, and so two DAC's were combined by a summing amplifier to obtain an effective 24-bit DAC as in Fig. \ref{dacout}, whose digitalization step (149 nV) is well below the analog voltage noise.  The requisite buffering and voltage protection was implemented on a printed circuit board, onto which the minikit was soldered via header pins.  In order to make use of all five ADCs, a potentiometer must be removed from the minikit.  A voltage reference for the ADCs and potentiometers is provided by the microcontroller itself.  Several digital general purpose input/output (GPIO) pins were dedicated to detecting user conditions such as scanning/lock mode option, and error signal sign inversion.  One output is used for scope blanking during the scan return after a rail condition occurs, to conceal hysteresis in the laser piezo during the return.

\begin{figure}
\includegraphics[width=6in]{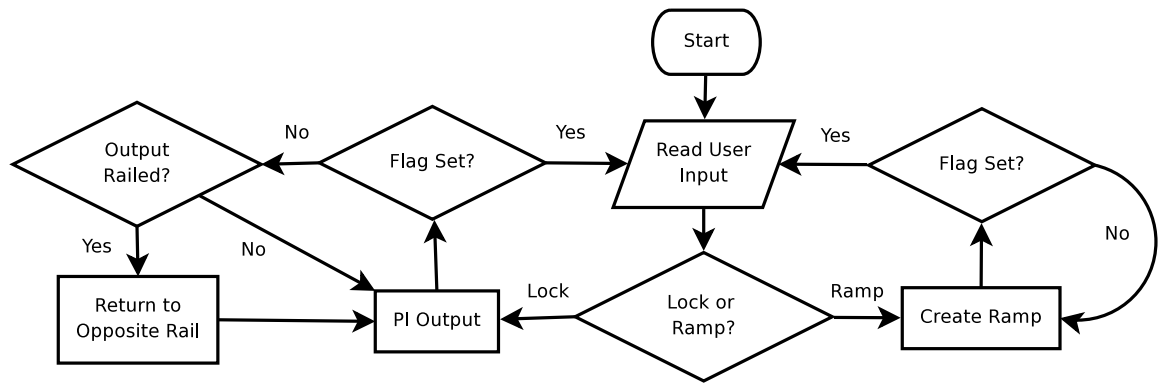}
\caption{A flowchart of the program execution.  All user settings, including gain fine tuning and mode switch, are detected during the read user input stage.  The flag mentioned is set by a timer interrupt which is triggered every 0.1 s.  The ADC is continuously converting in parallel with all steps during the lock branch of the diagram, but is read out and acted upon during the PI output stage.}
\label{flowchart}
\end{figure}

Program execution consists of two tightly written loops, one for sweeping and another for locking, as seen in Fig. \ref{flowchart}.  An internal timer is used to trigger an interrupt at the rate of about 10 Hz, which then sets a flag.  This flag indicates that the processor should spend a few clock cycles at the earliest convenience to read out the user controls in order to respond to any changes.  At this time the program determines the settings for the gain and sweep parameters, as well as which loop it should reenter, according to user setting.  The sweeping routine simply divides the set scanning range into a fixed number of steps and ramps over that range, pausing for a few microseconds between each increment.  Separate rail return functions operate in essentially the same way except with a larger step increment, making the ramp return faster than the sweep ramp.  The lock loop sets the ADC to continuous mode,  and then performs analysis on the previously acquired data point in order to achieve maximum parallelism.  If it detects that the calculated output has reached one of the user defined rails, it calls the aforementioned rail return function in order to restart the search.  Since the program always returns to the opposite rail that it previously encountered, it doesn't matter whether the error signal has a slight positive or negative offset, the PI will reacquire.

The microcontroller was programmed in C using the GNU compiler collection (GCC) version 4.1.1 from the ``WinARM'' package, and then uploaded using the serial cable and software sold with the minikit.  Since the ADuC7020 contains a common ARM7TDMI\textregistered~ processor, it can actually be programmed by a large number of compilers, both commercial and freely available.  Because the microcontroller has a slow 16-bit data bus to flash memory where program code is usually stored, we only obtained optimal performance by loading the code into SRAM at startup, which can be accomplished by assigning the lock loop function to the ``.data'' section with the appropriate compiler directive.

The sampling frequency was generally measured by adding a line of code that toggled one GPIO port during each execution cycle.  The sampling frequency is twice the measured frequency of that waveform.  Since this method increases the loop execution time slightly, its accuracy was cross checked by inputing a sine wave into the circuit and observing aliasing of the output directly.  This confirmed the accuracy of the port toggling method within the measurement accuracy of the later method, roughly 5\%.

Unfortunately the ARM7TDMI\textregistered~core contains no hardware floating point unit (FPU).  As a result, it is necessary to use a fixed-point scheme in order to effect the possibility of non-integer gain coefficients.  Our choice to combine two DACs led to a natural 24 bit signed fixed point data type, with the decimal point placed between the upper and lower 12 bits.  Division by factors not a power of 2 was avoided, so that all division operations could be performed by bit shift operations.  The two gains were in this way scaled to be between 0 and 1, with a switch option to make both gains negative.  This amounts to forcing the gains to occupy the 12 lowest bits of our invented data type.  The signal that the integral gain operates on during the feedback calculation stage, however, is divided down by $2^{10}$ from the actual measured integral of the error.  The error integral is simply the sum of all previous error measurements, which is reset to zero upon encountering a rail.  Dividing down the error integral avoids the need to divide the integral gain by a further $2^{10}$, which would result in only 4 possible settings.  Similarly, the error used in calculation of the proportional gain is divided by $2^5$, which has the added benefit of making the lock less sensitive to electronic noise.  Since the ADC reading has an offset subtracted from it, the error signal points range from -.5 to +.5, on the same scale.  On output, the upper 12 bits are written to the more strongly weighted DAC, and the lower 12 to the other.  Because of the rail detect function, negative values do not occur on output.

\section{Results and Discussion}

Before loading the lock loop into SRAM, the fastest lock speed was achieved by compiling the C code into ARM\textregistered~thumb mode commands, whereby we obtained a 250 kHz sampling rate, versus 200 kHz using the full complement of ARM\textregistered~commands.  Here, thumb mode is faster because each command is only 16 bits as opposed to 32 bits in full ARM\textregistered~mode, and since the flash memory bus is only 16 bits wide, 32 bit commands require two clock cycles to load.  The SRAM bus however is 32 bits, and so when loaded there the lock routine runs faster in full ARM\textregistered~mode, when it samples at a rate varying between 400 and 600 kHz.  This variation happens because the loop execution time is not commensurate with the ADC sampling time, and allowing the rate to vary does not seem to affect the lock's behavior.  Rewriting the routine in assembly could potentially allow for a consistently high rate, although now the bandwidth is limited by the overhead of checking if the output has railed or the flag has been set.

Aliasing of noise above the Nyquist frequency, especially near the sampling frequency, causes poor behavior of the feedback lock.  In some situations, aggressive active filtering would be applied to maximize bandwidth while eliminating any aliasing.  However, here using a filter with more than 2 poles will result in the deterioration of phase margin and in many cases oscillation of the feedback loop.  Therefore, for optimal noise suppression, the input filter's bandwidth should not exceed perhaps 40-60 kHz, depending on the filter design and system requirements\cite{Ellis}.

We have demonstrated the use of a microcontroller for direct frequency locking of  a fiber laser.  The same methods could be effectively applied to other lasers whose frequency can be quickly controlled by an external voltage, such as ECDLs.  Indeed the system is highly general, and could just as easily be used to close a doubling cavity, for example, with only software modification needed to course adjust the gain settings.  The electronic circuit itself is completely generic, and could be reprogrammed to address virtually any laboratory electronics problem that requires bandwidth on the 10-100 kHz scale.

The authors would like to acknowledge Adam Kleczewski and Chris Dostert for useful discussions leading up to this work.  This research was supported by the National Science Foundation Grant No. 0758025.


\end{document}